\documentclass[a4paper,twocolumn,floatfix]{revtex4}
\usepackage{graphicx}

\begin{document}

\title{Failed ``nonaccelerating" models of prokaryote gene
regulatory networks}

\author{M. J. Gagen and J. S. Mattick}

\affiliation{ARC Special Research Centre for Functional and
Applied Genomics, Institute for Molecular Bioscience, University
of Queensland, Brisbane, Qld 4072, Australia}

\email{m.gagen@imb.uq.edu.au}

\date{\today}

\begin{abstract}
Much current network analysis is predicated on the assumption that
important biological networks will either possess scale free or
exponential statistics which are independent of network size
allowing unconstrained network growth over time.  In this paper,
we demonstrate that such network growth models are unable to
explain recent comparative genomics results on the growth of
prokaryote regulatory gene networks as a function of gene number.
This failure largely results as prokaryote regulatory gene
networks are ``accelerating" and have total link numbers growing
faster than linearly with network size and so can exhibit
transitions from stationary to nonstationary statistics and from
random to scale-free to regular statistics at particular critical
network sizes. In the limit, these networks can undergo
transitions so marked as to constrain network sizes to be below
some critical value. This is of interest as the regulatory gene
networks of single celled prokaryotes are indeed characterized by
an accelerating quadratic growth with gene count and are size
constrained to be less than about 10,000 genes encoded in DNA
sequence of less than about 10 megabases. We develop two
``nonaccelerating" network models of prokaryote regulatory gene
networks in an endeavor to match observation and demonstrate that
these approaches fail to reproduce observed statistics.
\end{abstract}

\maketitle

\section{Introduction}

The difficulty of developing fully scalable technologies which can
be equally applied to both very small and very large systems
explains much of the current fascination with network analysis.
This field examines how growing networks can display stationary
(size independent) scale free or exponential statistics which are
unchanging over vast size ranges, and this field will naturally
focus on the very large and obvious networks possessing readily
obtainable statistics such as the Internet, the World Wide Web and
movie databases. However, there is an entire class of networks
equally important to human society, technology and biology which
possess nonstationary (size dependent) connectivity statistics and
which are thereby forced to undergo structural transitions as they
grow sometimes so severe as to limit growth entirely---for a
review see \cite{Gagen_03_accel_survey}.  The resulting limited
size of these networks makes them less obvious but does not
decrease their relevance.

In particular, prokaryote gene regulatory networks exploiting
homology based (sequence specific) interactions will display
nonstationary or ``accelerating" statistics where the link number
per node grows linearly with network size (so total link number
grows quadratically with network size), so these networks will be
inherently constrained to have sizes less than about 20,000 genes
\cite{Croft_03_unpub}. In fact, all prokaryotic gene numbers and
genomes are indeed of restricted size (less than about 10,000
genes with genomes of between 0.5 and 10 megabases
\cite{Casjens_98_33}), in contrast to the genomes of multicellular
eukaryotes (with for humans, about 30,000 genes and a genome of
about 3 gigabases \cite{Int_Human_genome_01_86,Venter_01_13}).

The rapidly expanding field of network analysis, reviewed in
\cite{Dorogovtsev_02_10,Albert_02_47}, has provided examples of
networks exhibiting ``accelerating" network growth where link
number grows faster than linearly with network size
\cite{Dorogovtsev_01_025101,Sen_0310513}.  For instance, the
Internet \cite{Faloutsos_99_25} appears to grow by adding links
more quickly than sites though the relative change over time is
small and the Internet appears to remain scale free and well
characterized by stationary statistics \cite{Vasquez_02_066130}.
Similarly, the number of links per substrate in the metabolic
networks of organisms appears to increase linearly with substrate
number \cite{Jeong_00_65}, the average number of links per
scientist in collaboration networks increases linearly over time
\cite{Dorogovtsev_00_33,Vasquez_00_0006132,Barabasi_01_0104162,Barabasi_02_590,Vasquez_03_056104},
and languages appear to evolve via accelerated growth
\cite{Dorogovtsev_01_2603}.  Even social networks take on their
small world characteristics only when the network is large
enough---in small towns everyone knows everyone else so social
networks are accelerating and exhibit a transition to small world
statistics only as individual nodes saturate their connectivity
limits \cite{Watts_99_493}. Accelerating networks are more
prevalent and important in society and in biology than is commonly
realized \cite{Gagen_03_accel_survey}.

A ``probabilistic" accelerating model of prokaryote regulatory
gene networks has been developed in Ref. \cite{Gagen_0312021}.
This involved the use of probabilistic links to allow arbitrarily
rapid acceleration rates, two distinct classes of nodes where
``regulators" can source outbound regulatory links to regulate
other nodes (both regulators and non-regulators) while
``non-regulators" cannot source outbound links, directed links
from regulators to regulated nodes, and distinct connectivity
distributions describing the long-tailed and scale-free
distribution of outbound link number per regulator and the compact
and exponential distribution of the inbound link number per node.
The resulting model satisfactorily matched observable parameters.
However, this success is meaningless if similar results can be
achieved via nonaccelerating network models.  In this paper, we
will show that the two simplest nonaccelerating network models
fail to explain either the observed quadratic growth of regulator
number with genome size or the detailed statistics pertaining to
the {\em E. coli} genome.

In Section \ref{sect_overview_prok_networks} we canvass the
available literature to characterize the statistics of prokaryote
gene regulatory networks.  This then allows the construction of
two nonaccelerating network models in Section
\ref{sect_non_accel_model} where we use the continuous
approximation and simulations to analyze network statistics
allowing comparison to observation.

\begin{figure}[htbp]
\centering
\includegraphics[width=0.9\columnwidth,clip]{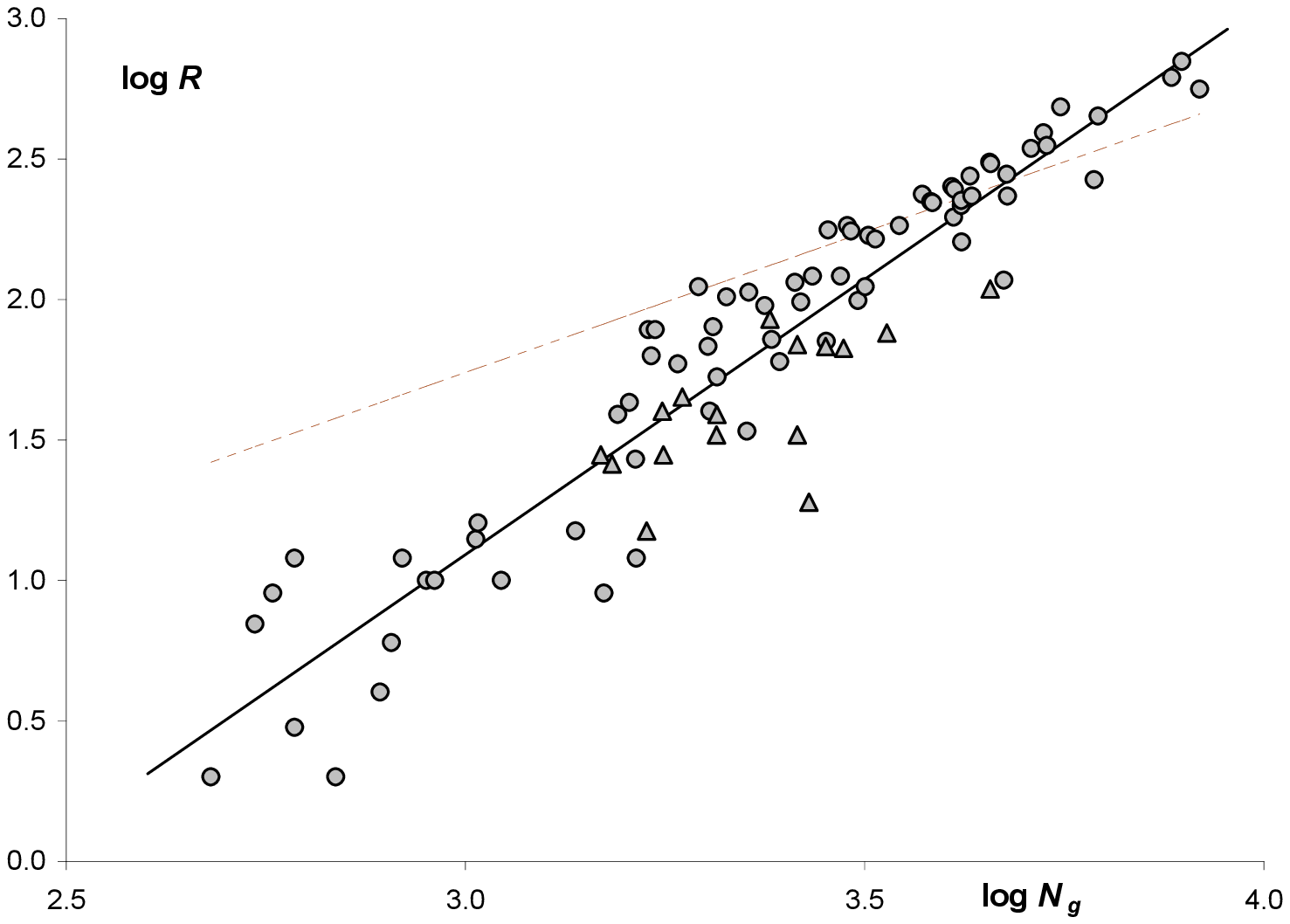}%
\hspace{-4cm}
\includegraphics[width=0.5\columnwidth]{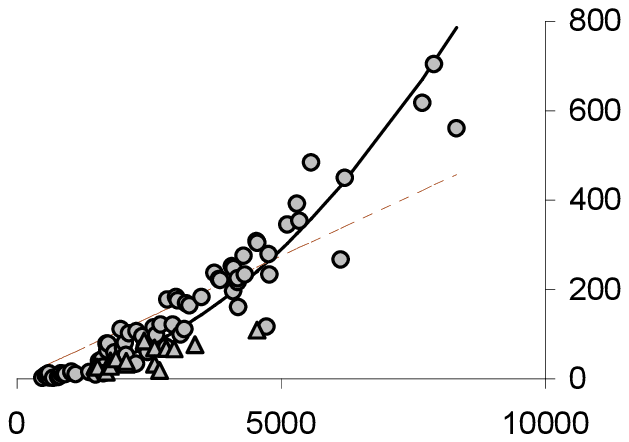}
\caption{\em Double-logarithmic plot of regulatory protein number
($R$) against total gene number ($N_g$) for bacteria (circles) and
archaea (triangles), adapted from Ref.
\protect\cite{Croft_03_unpub}.  The log-log distribution is well
described by a straight line with slope $1.96\pm 0.15$ ($r^2=
0.88$, 95\% confidence interval indicated), corresponding to a
quadratic relationship between regulator number and genome size.
The inset shows the same data before log-transformation
\protect\cite{Croft_03_unpub}. Dashed lines show the best linear
fit to the data.} \label{f_prok_reg_data}
\end{figure}

\section{Overview of prokaryote gene networks}
\label{sect_overview_prok_networks}

Ongoing genome projects are now providing sufficient data to
usefully constrain analysis of the gene regulatory networks of the
simpler organisms. Ref. \cite{vanNimwegen_03_479} first noted
quadratic growth in the class of transcriptional regulators ($R$)
with the number of genes $N_g$ in bacteria with the observed
results
\begin{equation}
   R\propto \left\{
     \begin{array}{ll}
       N_g^{1.87\pm0.13}, & \mbox{transcriptional regulation} \\
        & \\
       N_g^{2.07\pm0.21}, & \mbox{two component systems} \\
        & \\
       N_g^{2.03\pm0.13}, & \mbox{transcriptional regulation} \\
        & \\
       N_g^{2.16\pm0.26}, & \mbox{transcriptional regulation}. \\
     \end{array}
    \right.
\end{equation}
Here, the top two lines refer to different classes of regulators
while the bottom two lines are the results of a crosschecking
analysis of two alternate databases, and quoted intervals reflect
99\% confidence limits \cite{vanNimwegen_03_479}. Ref.
\cite{Croft_03_unpub} analyzed of 89 bacterial and archeael
genomes to determine the relations
\begin{equation}     \label{eq_Croft_results}
   R= \left\{
     \begin{array}{ll}
       aN_g^b=(1.6\pm0.8)10^{-5}N_g^{1.96\pm0.15} & (r^2=0.88) \\
        &  \\
       pN_g^2=(1.10\pm0.06)10^{-5}N_g^{2} & (r^2=0.87) \\
        &  \\
       cN_g=(0.055\pm0.004)N_g  & (r^2=0.75). \\
     \end{array}
    \right.
\end{equation}
In all cases, the limits reflect 95\% confidence levels, and for
completeness, the data is shown in Fig. \ref{f_prok_reg_data}. The
observed quadratic growth implies an ever growing regulatory
overhead so there will eventually come a point where continued
genome growth requires the number of new regulators to exceed the
number of nonregulatory nodes, and based on this, Ref.
\cite{Croft_03_unpub} predicted an upper size limit of about
20,000 genes, within a factor of two of the observed ceiling. A
number of other papers have noted the faster than linear growth of
regulator number with genome size. In particular, it was noted
that larger genomes harboured more transcription factors per gene
than smaller ones \cite{Cases_03_248}, and that regulators form an
increasing proportion of all genes as genome size increases
\cite{Stover_00_959,Bentley_02_141}.

Prokaryotes typically group their DNA encoded genes in operons,
co-regulated functional modules of average size 1.70 genes each in
{\em E. coli} which value we treat as typical though in reality,
operon size decreases slightly with genome size
\cite{Cherry_03_40}. {\em E. coli} regulatory proteins affect an
average of about 5 operons with this distribution being long
tailed \cite{Shen_Orr_02_64} so the majority of regulators affect
only one operon though some regulators (CRP) can affect up to 71
operons or 133 genes \cite{Thieffry_98_43}.  More recent estimates
show this transcription factor---CRP, a global sensor of food
levels in the environment---regulating up to 197 genes directly
and a further 113 genes indirectly via 18 other transcription
factors \cite{MandanBabu_03_1234}.  (To observe the long tailed
distribution, see Fig. 2 of Ref. \cite{Thieffry_98_43} and Fig. 4
of Ref. \cite{MandanBabu_03_1234}.)

The number of inputs taken by an operon is characterized by a
compact exponential distribution with a rapidly decaying tail so
the majority of regulated operons are controlled by a single
regulator while very few regulated operons are controlled by four,
five, six or seven regulators
\cite{Thieffry_98_43,Shen_Orr_02_64,MandanBabu_03_1234}. The
average number of inputs in {\em E. coli} is about 1.4
\cite{Shen_Orr_02_64}, 1.5 \cite{Thieffry_98_43}, or 1.6.
\cite{MandanBabu_03_1234}.  In addition, 31.4\% of {\em E. coli}
transcription factors regulate other transcription factors
\cite{MandanBabu_03_1234}, while 37.7\% of non-autoregulatory
cascades in {\em E. coli} are of length two, 52.5\% are
three-level cascades, and 9.8\% are four-level cascades
\cite{MandanBabu_03_1234}.

\section{Nonaccelerating prokaryote network models}
\label{sect_non_accel_model}

We extend the gene network model of Refs.
\cite{Thieffry_98_43,Gagen_0312021} to construct two
nonaccelerating network models of prokaryote regulatory gene
networks. Prokaryotes typically pack their $N_g$ genes into a
lesser number of $N=N_g/g_o$ co-regulated operons where we assume
that operons contain exactly $g_o=1.70$ genes. Of the existing
operons, $O_r$ are regulated operons and $O_u=N-O_r$ are
unregulated operons.  Of the total number of operons, there are
$R$ regulatory operons whose regulatory interactions are directed
links from regulatory operons to regulated operons. Under the
assumption that there is only one regulatory gene per regulatory
operon, the observed linear relation of Eq. \ref{eq_Croft_results}
becomes
\begin{equation}   \label{eq_reg_No}
    R=cN_g \; = \; c g_o N.
\end{equation}
In nonaccelerating network models, the number of links per
regulator is constant so consequently, the total number of links
must increase linearly with network size, giving
\begin{equation}   \label{eq_link_No}
    L=lN.
\end{equation}
Here, the value for $l$ will be approximately $cg_o=0.0935$, but
the exact relation must be derived from the details of the
implemented model.

\begin{figure}[htb]
\centering
\includegraphics[width=\columnwidth,clip]{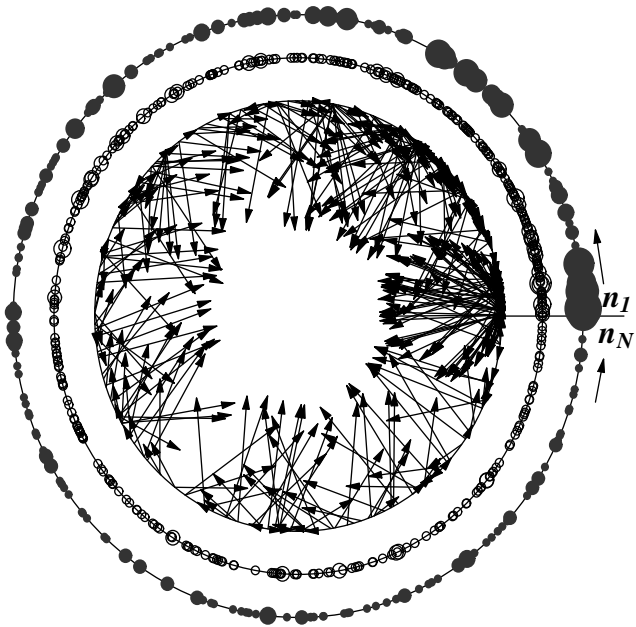}
\caption{\em An example statistically generated E. coli genome
using the later results of the one-parameter model, where (for
convenience only) operon nodes numbered $n_1, \dots, n_N$ are
placed sequentially counterclockwise on a circle in their
historical order of entry into the genome. The filled points on
the outer circle locate regulators and have radius indicating the
number of outbound regulatory links. The open points on the middle
circle locate regulated operons and have radius indicating the
number of inbound regulatory inputs.  The arrows in the inner
circle show all directed regulatory
links.}\label{f_e_coli_constant_p1}
\end{figure}

Following Ref. \cite{Gagen_0312021}, each regulatory link between
nodes is directed, and characterized by two distinct distributions
describing respectively the placement of the heads and tails of
each link. Only a relatively few nodes are regulatory, and of
these, the number of outbound link tails per regulatory node are
described by a size dependent long-tailed distribution with
average about $\langle t\rangle\approx 5$. Such a long-tailed
distribution requires that link tails be preferentially attached
to an existing regulatory operon, and this requirement places
restrictions on the gene duplication processes assumed by the
model---see Ref. \cite{Gagen_0312021} for details. In contrast to
the relatively small number of regulatory nodes, all nodes can
themselves be multiply regulated by inbound links.  Further, the
many used and unused promotor region binding sites broadly sample
the space of possible binding sites so only a small fraction of
nodes will be regulated by any one regulator. As a result, the
number of inbound link heads per node is described by a size
dependent exponential distribution with a low average of $\langle
h\rangle\approx 1.5$ as typically results from the random or
non-preferential attachment of inbound links to operon promotor
sequences.

We suppose that the operon network grows by the sequential
addition of numbered nodes $n_k$ for $1\leq k\leq N$, and that at
network size $k$, node $n_i$ ($1\leq i\leq k$) has $t_{ik}$
outbound tails and $h_{ik}$ inbound heads.  We do not model the
many trials of potential genes over many generations and merely
include fixated genes in our count---that is, drifting sequence is
not counted as part of the fixated genome.  This further implies
the sequence of established nodes is under severe selective
constraint and unable to drift so consequently new links cannot be
added between existing nodes.

For clarity, Fig. \ref{f_e_coli_constant_p1} preempts later
calculations (from the one-parameter model) and depicts a
statistically generated version of an {\em E. coli} genome where
nodes are placed sequentially counterclockwise in a circle (for
convenience only). Alternative genome models may be distinguished
by the age distribution of regulators, regulated operons and their
link numbers, and these are indicated in this figure. In
particular, Fig. \ref{f_e_coli_constant_p1} shows a highly
nonuniform distribution of both regulators and outbound link
numbers and of regulated operons and inbound link numbers with
gene age. (These age-independent distributions are in marked
contrast to those generated by accelerating models of regulatory
gene networks \cite{Gagen_0312021}.)

A substantial proportion of the gene regulation network of
prokaryotes is enacted via homology dependent interactions as when
sequence specified protein transcription factors bind to specific
promoter sequences. Naturally then, regulators will form more
links in larger genomes than in smaller genomes
\cite{Croft_03_unpub,Gagen_0312021}.  Such interactions lead
immediately to accelerating models of gene regulatory networks
\cite{Gagen_0312021}, making it difficult to propose plausible
physical mechanisms restricting regulators to form the same
probable number of initial links independent of genome size and
thereby implement a nonaccelerating model. However, the purpose of
this paper is to fully evaluate nonaccelerating gene regulatory
network models, and we here presume that such physical mechanisms
exist (without detailing them).

\subsection{One-parameter model}

For our first model, we assume that on entry into the genome, each
new node $n_k$ can form a total of up to $m$ outbound regulatory
links with all the nodes $n_1, \dots, n_k$ with each individual
link forming with probability $p$, and, provided that sufficient
regulators already exist, up to a total of $m$ inbound regulatory
links each with probability $p$ from some subset of the existing
regulators chosen according to preferential attachment. (For
consistency, the probable number of inbound distinct regulatory
links to node $n_k$ ($\approx mp$) must be less than the probable
number of existing regulators ($cg_oN$), satisfied when genomes
have size $N>mp/cg_o\approx 1$.) Hence, the respective
probabilities that the initial number of heads $h_{kk}=j$ or the
initial number of tails $t_{kk}=j$ for node $n_k$ is
\begin{equation}   \label{eq_P_j_k_dist}
 P(j) = {m \choose j} p^j (1-p)^{m-j},
\end{equation}
with the proviso that all the inbound links can only be added to
node $n_k$ if there is a sufficient number of regulators among the
nodes $n_1, \dots, n_k$. The average number of inbound and
outbound links is identical, $\langle t_{kk}\rangle=\langle
h_{kk}\rangle=mp$ independent of network size. The addition of
node $n_k$ and its links will increase the probable number of
heads attached to earlier nodes $n_j$ for $1\leq j\leq (k-1)$ so
$h_{jk}\geq h_{jj}$, while the probable number of tails outbound
from node $n_j$ increases $t_{jk}\geq t_{jj}$ if and only if that
node is regulatory with $t_{jj}>0$.

The average number of links in a network of size $N$ nodes is then
\begin{equation}          \label{eq_total_links}
   L   = 2mpN \;=\; l N,
\end{equation}
taking account of both heads and tails. Under the assumption that
regulators can only be created on entry to the genome
\cite{Gagen_0312021}, the distribution of regulators at any time
is specified by the distribution $P(j)$ for $t_{kk}$ so the
probability that node $n_k$ is a regulator is $1-P(0,k)$.  For a
network of $N$ nodes, the predicted total number of regulators is
then
\begin{eqnarray}      \label{eq_reg_density}
  R &=&  \sum_{k=1}^{N} \left[ 1- (1-p)^m \right]  \nonumber  \\
    &=&  \left[ 1- (1-p)^m \right] N \;=\; cg_oN.
\end{eqnarray}
The bottom line shows the expected behaviour for the number of
regulators in the respective limits $p\rightarrow 0$ giving
$R\rightarrow 0$, and $p\rightarrow 1$ giving $R\rightarrow N$.
Comparison to the observed Eq. \ref{eq_reg_No} provides the noted
constraint which reduces the number of free variables by one to
justify this as a one-parameter model. Combining Eqs.
\ref{eq_reg_No}, \ref{eq_total_links} and \ref{eq_reg_density},
and noting that $m$ is integral gives
\begin{equation}
   l = 2mp\;=\;2m \left[1- (1-cg_o)^{1/m} \right], \; m=1,2,\dots,
\end{equation}
which establishes the infinite number of possible modelling
choices
\begin{eqnarray}      \label{eq_constraint_family}
  (m,p,l) &=&  (1,0.0935,0.187) \nonumber  \\
   &=&  (20,0.00490,0.196) \nonumber  \\
   &=&  (40,0.00245,0.196) \nonumber \\
   &&  \vdots
\end{eqnarray}
The values of the link formation probability $p$ over this range
of $m$ values suggest overly short average promotor binding site
lengths of between $-\log_4 p\in[1.7,4.3]$ bases. These values are
unreasonably low though we are restricted from exploring
arbitrarily large values for $m$ by our desire to develop a
nonaccelerating network model---obtaining a promotor sequence
length of about 6 requires $m>400$, and such large $m$ values
effectively implement an accelerating network model as every
regulator can effectively explore links to every operon in even
large genomes. For modelling purposes, we set $m=20$ and
$p=0.00490$ to give
\begin{equation}                \label{eq_l_assignment}
   l = 0.196.
\end{equation}
This high link formation rate leads to the heavy density of
regulators and regulated operons in Fig.
\ref{f_e_coli_constant_p1}. The average number of links per
regulator using Eqs. \ref{eq_reg_density} and
\ref{eq_l_assignment} is then approximately $L/R=l/(cg_o)=2.10$, a
constant for all genomes which is reasonably close to the observed
value of $5$ for {\em E. coli} \cite{Shen_Orr_02_64}.

\subsubsection{Random distribution of regulated operons---I}

The distribution of link heads for all nodes (with possession of a
link head designating a regulated node), can be straightforwardly
calculated under the assumption that the $t_{kk}\approx mp=l/2$
new tails added with node $n_k$ are randomly distributed across
the $k$ existing nodes so on average, each existing node receives
$l/2k$ additional inbound links. The continuous approximation
\cite{Barabasi_99_17,Barabasi_99_50,Dorogovtsev_01_056125} for
links randomly distributed over $k$ existing nodes determines the
number of inbound head links for node $n_j$ according to
\begin{equation}        \label{eq_continuum_heads}
  \frac{\partial h_{jk}}{\partial k}= \frac{t_{kk}}{k}
   \; = \; \frac{l}{2k}.
\end{equation}
This can be integrated with initial conditions $h_{jj}\approx l/2$
at time $j$ and final conditions $t_{jN}\approx l/2$ at time $N$
to give
\begin{equation}          \label{eq_hjn}
  h_{jN} = \frac{l}{2} \left[1+\ln\left(\frac{N}{j} \right)\right].
\end{equation}
The number of inbound regulatory links per node is then dependent
on the age of each node. Integration of these link numbers over
all node numbers $j$ gives the required total number of links as
in Eq. \ref{eq_total_links}.  This distribution suggests that the
oldest node $n_1$ for the {\em E. coli} genome with $N=2528$ nodes
will possess an average of $h_{1N}=0.87$ inbound regulatory links
while the most recent node $n_{NN}$ will possess an average of
$h_{NN}=0.098$ inbound regulatory links---see the age dependent
distributions of Fig. \ref{f_e_coli_constant_p1}.

The very useful continuum approach is not entirely accurate when
applied to these nonaccelerating networks, and it is necessary to
check later results using fuller derivations of the underlying
joint probability distributions.  In particular, the probability
that by time $N$, node $n_k$ has received an initial
$h_{kk}=j\in\{0,m\}$ inbound links each with probability $p$, and
subsequently received $j_k\in\{0,m\}$ inbound links from itself
each with probability $p/k$, as well as $j_{k+1}\in\{0,m\}$
inbound links from node $n_{k+1}$ each with probability $p/(k+1)$,
and so on until it receives $j_N\in\{0,m\}$ inbound links from
node $n_N$ each with probability $p/N$, is
\begin{eqnarray}              \label{eq_pjjkjk1_etc}
   P(j,j_k,j_{k+1},\dots,j_N) &=&  \nonumber \\
   && \hspace{-4.5cm} {m \choose j} p^j (1-p)^{m-j}
   \prod_{n=k}^N
     {m \choose j_n} \left(\frac{p}{n}\right)^{j_n}
     \left[1-\frac{p}{n} \right]^{m-j_n}.
\end{eqnarray}
The average number of inbound links for node $n_k$ is then
\begin{eqnarray}
   \langle j+j_k+\dots+j_N\rangle
   &=&\frac{l}{2}(1+1/k+\dots+1/N) \nonumber \\
   &\approx& \frac{l}{2}\left[1+\ln\left(\frac{N}{j} \right)\right]
\end{eqnarray}
as found by the continuum approach. (Later results will not match
so closely.)

The number of links per node is monotonically decreasing with node
number as even though all nodes receive the same number of initial
links on average, earlier nodes have a longer time to accumulate
more links than later nodes. This distribution contains
information about both node connectivity and node age and so
approximates genome statistics (simulated or observed) when all
this information is available. However, it is usually the case
that node age information is unavailable necessitating calculation
of connectivity distributions that are not conditioned on node
age. This effectively requires binning together all nodes
irrespective of their age to obtain a final link distribution. We
can use the continuum approach for monotonically decreasing link
numbers with node age
\cite{Barabasi_99_17,Barabasi_99_50,Dorogovtsev_01_056125} to
discard the often unknown age information via
\begin{eqnarray}         \label{eq_final_link_dist}
  H(k,N) &=& \frac{1}{N} \int_0^N dj \; \delta(k-h_{jN})\nonumber \\
  & = & - \frac{1}{N} \left( \frac{\partial h_{jN}}{\partial j}
  \right)^{-1}  \mbox{at }[j=j(k,N)],
\end{eqnarray}
where $j(k,N)$ is the solution of the equation $k=h_{jN}$. For our
case with the constraint $k=l(1+\ln(N/j))/2$, the final
distribution of link heads absent age information is
\begin{equation}                 \label{eq_hkn_continuous}
   H(k,N) = \frac{2}{l} e^{\left( 1-2k/l\right)},
\end{equation}
showing an exponentially rapid decrease in the number of probable
links.  As every node initially receives a minimum of $l/2$ links,
this distribution is normalized, $\int_{l/2}^{\infty} H(k,N)=1$,
and has average $\langle k \rangle=\int_{l/2}^{\infty} kH(k,N)=l$.
The expected proportion of nodes $P_h(k)$ possessing $k$ inbound
links is then obtained by integrating the continuous distribution
of Eq. \ref{eq_hkn_continuous} over appropriate ranges $[l/2,1/2]$
or $[k-1/2,k+1/2]$ to obtain
\begin{equation}             \label{eq_all_link_number}
    P_h(k) = \left\{
      \begin{array}{ll}
        1 - e^{\left( 1-1/l\right)}, & k=0 \\
         &  \\
        2 \sinh \left(1/l\right) e^{\left( 1-2k/l\right)}, & k>0. \\
      \end{array}
    \right.
\end{equation}
Consequently, the distribution of inbound link numbers for
regulated nodes (i.e. those with $k>0$) is $P_h(k)/[1-P_h(0)]$, or
\begin{equation}            \label{eq_reg_link_number}
  P_r(k) =  \left( e^{2/l}-1 \right) e^{-2k/l},
\end{equation}
which again is normalized to unity.

\begin{figure}[htbp]
\centering
\includegraphics[width=\columnwidth,clip]{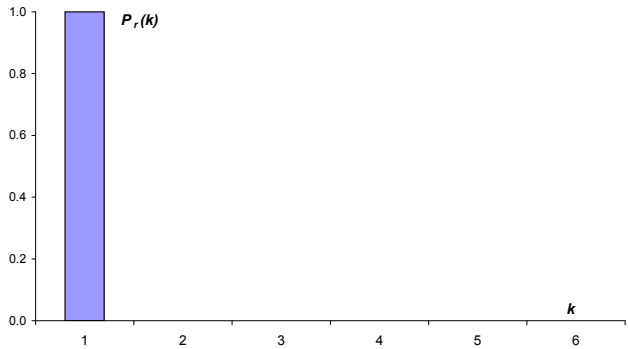}
\caption{\em The predicted proportions $P_r(k)$ of the regulated
operons of E. coli taking multiple regulatory inputs for genomes
of any size. This distribution poorly approximates that observed
for E. coli with $N=2528$ operons in Fig. 2(d) of Ref.
\protect\cite{Cherry_03_40} and of Fig. 5 of Ref.
\protect\cite{MandanBabu_03_1234}. \label{f_heads_per_reg_operon}}
\end{figure}

These distributions for the number of inbound link heads per node
and per regulated node permit the calculation of the number of
unregulated operons $O_u$ via either $P(0,\dots,0)$ (from Eq.
\ref{eq_pjjkjk1_etc}) or $P_h(0)$ (from Eq.
\ref{eq_all_link_number}).  Thus, the total number of unregulated
nodes is respectively
\begin{equation}      \label{eq_unreg_operons}
   O_u= \left\{
      \begin{array}{l}
        \sum_{k=1}^N (1-p)^m
      \prod_{n=k}^N \left[1-\frac{p}{n} \right]^{m} \\
        \\
       N\left[ 1 - e^{\left( 1-1/l\right)}\right]. \\
      \end{array}
     \right.
\end{equation}
The top line here shows the expected behaviour with $p\rightarrow
1$ giving $O_u\rightarrow 0$ and $p\rightarrow 0$ giving
$O_u\rightarrow N-lN$ as required. The second line derived using
the continuum approximation fails to exhibit the desired
dependency on link number as $l\rightarrow 0$ demonstrating that
care must be taken in using this approach. Using the more accurate
top line, the number of regulated nodes is then approximately
$O_r=N-O_u\approx lN$, so in turn, the number of inbound links per
regulated node is $L/O_r=1$.  A direct calculation of the average
number of inbound links for regulated operons using the
distribution of Eq. \ref{eq_reg_link_number} gives $\langle
k\rangle=\sum_{k=1}^{\infty} kP_r(k)=1/[1-\exp(-2/l)]=1.00004$,
close to the value of $1.5$ or 1.6 observed in {\em E. coli}
\cite{Thieffry_98_43,Shen_Orr_02_64,MandanBabu_03_1234}.  In
addition, the average number of inbound regulatory links per
operon (for all operons) is $\langle k\rangle=L/N=l=0.196$. The
predicted distribution of inbound links for regulated operons (Eq.
\ref{eq_reg_link_number}) can be compared to that observed in the
{\em E. coli} network of size $N=2528$ operons
\cite{Cherry_03_40}, and is shown in Fig.
\ref{f_heads_per_reg_operon}. The overly rapid decay of the
calculated distribution poorly approximates the compact
exponential distribution observed for {\em E. coli} shown in Fig.
2(d) of Ref. \cite{Cherry_03_40} and of Fig. 5 of Ref.
\cite{MandanBabu_03_1234} leading to an underestimation of the
numbers of regulated operons with 2 or more inputs---essentially
no regulators are predicted to have 2 or more inputs for genomes
of size $N=2528$ operons.

\subsubsection{Scale-free distribution of regulator operons---I}

At time $k$, the $h_{kk}\approx l/2$ inbound links associated with
node $n_k$ have their tails preferentially attached to existing
regulatory nodes $n_j$ with probability proportional to the number
of existing regulatory links for that node at time $k$, i.e.
$t_{jk}$. Using the continuous approximation
\cite{Barabasi_99_17,Barabasi_99_50,Dorogovtsev_01_056125}, the
rate of growth in outbound link number for node $n_j$ is then
approximately
\begin{equation}    \label{eq_continuum_tails}
  \frac{\partial t_{jk}}{\partial k} =
     h_{kk}\frac{t_{jk}}{\int_{0}^{k} t_{jk} \; dj}.
\end{equation}
The denominator here is a probability weighting to ensure
normalization and is the total number of outbound links for all
nodes at network size $k$. Following \cite{Dorogovtsev_02_10}, we
can evaluate the denominator using the identity
\begin{equation}
  \frac{\partial}{\partial k} \int_0^k t_{jk} \;dj =
     \int_0^k  \frac{\partial}{\partial k} t_{jk} \; dj + t_{kk}.
\end{equation}
This can be evaluated using Eq. \ref{eq_continuum_tails} noting
$t_{kk}\approx h_{kk}\approx l/2$ giving
\begin{equation}
  \frac{\partial}{\partial k} \int_0^k t_{jk} \;dj = l,
\end{equation}
which can be integrated determining the denominator of Eq.
\ref{eq_continuum_tails} to be
\begin{equation}
   \int_0^k t_{jk} \;dj = l k.
\end{equation}
This is in agreement with Eq. \ref{eq_total_links}. Substituting
this value into Eq. \ref{eq_continuum_tails} gives
\begin{equation}
  \frac{\partial t_{jk}}{\partial k} =\frac{t_{jk}}{2k}.
\end{equation}
Finally, this can be integrated with initial conditions
$t_{jj}\approx l/2$ at time $j$ and final conditions $t_{jN}$ at
time $N$ to give
\begin{equation}     \label{eq_tails_t_jn}
  t_{jN} = \frac{l}{2}  \left(\frac{N}{j} \right)^{\frac{1}{2}}.
\end{equation}
Because we are now considering outbound links, we must take
account of our use of two classes of distinguishable nodes,
regulators and non-regulators, by allowing for the known
distribution of regulators with node number over the genome. The
average link number per node at node $n_j$ (Eq.
\ref{eq_tails_t_jn}) equates to the product of the average number
of link tails per regulator at node $n_j$, denoted $t_r(j,N)$, and
the average number of regulators per node at node $n_j$, denoted
$\rho(j)$.  This latter density is $\rho(j)=dR(j)/dj=cg_o$ by Eq.
\ref{eq_reg_density}, so by definition, we have
\begin{equation}
  t_{jN}= t_r(j,N) \rho(j),
\end{equation}
giving
\begin{equation}   \label{eq_t_ii_dist}
  t_r(j,N)= \frac{l}{2cg_o} \left(\frac{N}{j} \right)^{\frac{1}{2}}.
\end{equation}
Again we find a monotonically decreasing number of links per
regulator with node number or age so older nodes are more heavily
connected---see Fig. \ref{f_e_coli_constant_p1}. Our treatment
here effectively duplicates previous results for networks adding a
constant deterministic number of links per node
\cite{Albert_02_47}.

As usual, we again use the continuum approach for monotonically
decreasing link numbers with node age (Eq.
\ref{eq_final_link_dist} and noting $k=l(N/j)^{(1/2)}/2cg_o$)
\cite{Barabasi_99_17,Barabasi_99_50,Dorogovtsev_01_056125} to
discard the often unknown age information in the $t_r(j,N)$
distribution to obtain the outbound link distribution
\begin{equation}   \label{eq_k_minus_three}
   T(k,N) = \frac{1}{2} \left(\frac{l}{cg_o} \right)^2 \frac{1}{k^3},
\end{equation}
which is normalized over the range $[k_0=l/2cg_o\approx
1.05,\infty)$, as $\int_{k_0}^\infty T(k,N)=1$. In turn, the
expected proportion of regulators $P_t(k)$ possessing $k$ links is
then obtained by integrating the continuous distribution of Eq.
\ref{eq_k_minus_three} over appropriate ranges $[k_0,3/2]$ or
$[k-1/2,k+1/2]$ to obtain
\begin{equation}
    P_t(k) = \left\{
      \begin{array}{ll}
        1-\left( \frac{l}{3cg_o} \right)^2, & k=1 \\
         &  \\
        8 \left( \frac{l}{cg_o} \right)^2 \frac{k}{(4k^2-1)^2}, & k>1. \\
      \end{array}
    \right.
\end{equation}
As required, this is normalized to unity. The average number of
outbound links per regulator is, using the continuous distribution
$T(k,N)$, $\langle k\rangle=\int_{k_0}^\infty kT(k,N)=l/cg_o=2.09$
and numerically calculated to be $\langle k\rangle=1.98$ using the
$P_t(k)$ distribution (complementing previous estimates following
Eq. \ref{eq_l_assignment}) each of which compares well to the
observed value of 5 in {\em E. coli} \cite{Shen_Orr_02_64}.

\begin{figure}[htb]
\centering
\includegraphics[width=\columnwidth,clip]{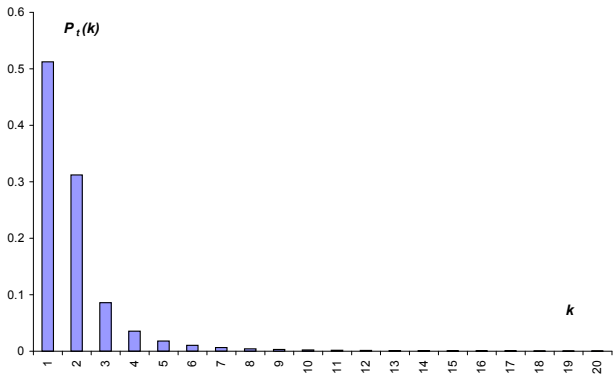}
\caption{\em The predicted proportion of regulatory operons
$P_t(k)$ regulating $k$ different operons for a simulated E. coli
genome with $N=2528$ operons.   As expected, most regulators
regulate only one other operon, though a small number of
regulators can regulate more than 10 operons. This distribution
poorly approximates the observed proportions for E. coli in Fig.
2(c) of Ref. \protect\cite{Cherry_03_40} and Fig. 4 of Ref.
\protect\cite{MandanBabu_03_1234}, and predicts the probable
existence of only one E. coli regulator possessing link numbers in
the range between $[20,\infty]$ links.
\label{f_e_coli_long_tails}}
\end{figure}

However, the very rapid (cubic) decrease in probable link numbers
means that these distributions have difficulty in reproducing the
distributions observed in {\em E. coli}. The expected outbound
link distribution appears in Fig. \ref{f_e_coli_long_tails}
showing a long-tailed and scale free distribution with
probabilities scaling roughly as $P_t(k)\propto k^{-3}$. The
$P_t(k)$ distribution shows that a full 51\% of regulators have
only one link, while 83\% have two or fewer links, and 91\% have
three or fewer links. In particular, the expected number of
regulators with $k$ links is $P_t(k)R$ with the number of
regulators $R$ obtained from Eq. \ref{eq_reg_density} (or from
observation). For {\em E. coli} with $N=2528$ operons
\cite{Cherry_03_40}, this predicts the probable existence of only
one {\em E. coli} regulator possessing link numbers in the range
between $[20,\infty]$ links. This poorly approximates the
connectivity of {\em E. coli} where many regulators regulate more
than 20 operons including the global food sensor CRP which
regulates up to 197 genes directly \cite{MandanBabu_03_1234}. In
fact, Eq. \ref{eq_t_ii_dist} with $j=1$ predicts that the most
heavily connected node in {\em E. coli} has only about $53$
downstream links.

\subsubsection{Cascades and regulatory islands---I}
\label{sect_size_constraints}

Nonaccelerating networks have stationary statistics which are
independent of network size. In particular, the proportion of the
network present in islands of nodes of various sizes connected by
regulatory links is independent of genome size. As prokaryote
regulatory networks likely consist of functionally distinct
regulated modules \cite{Thieffry_98_43,Hartwell_99_c4} with a
marked absence of regulatory cycles (at least in {\em E. coli}
\cite{Thieffry_98_43,Shen_Orr_02_64,MandanBabu_03_1234}), any
network model must be able to adequately reproduce the statistics
of regulatory islands and cascades.

The proportion of transcription factors which control downstream
regulators is
\begin{eqnarray}    \label{eq_regulation_regulators}
   P_{rr}(N) &=& \frac{1}{R} \sum_{k=1}^N
     \left[1-(1-p)^m \right]
     \left[ \frac{l}{2cg_o} \left(
     \frac{N}{k}\right)^{\frac{1}{2}}\right]
     \frac{R}{N} \nonumber  \\
     &\approx & l.
\end{eqnarray}
Here, the first fraction on the RHS normalizes the proportion in
terms of the number of regulators $R$ (Eq. \ref{eq_reg_density}),
the first term in the summation is the probability that node $n_k$
is a regulator, the second term is the average number of
regulatory outbound links for this regulatory node $t_r(k,N)$ at
network size $N$ (Eq. \ref{eq_t_ii_dist}), and the third term
approximates the probability that these nodes link to one of the
existing regulators under random attachment.  (If the very first
and very last terms are dropped, the remaining summation over all
nodes of the probability that $n_k$ is regulatory with the stated
number of links equates to the total number of links in the
network $L=lN$. This is the more accurate version of the
calculation leading to Eq. \ref{eq_t_ii_dist}.) Hence, the
proportion of regulators which control transcription factors is
independent of network size and equals 19.6\%.  This ratio
compares reasonably well with that observed in {\em E. coli} where
Ref. \cite{MandanBabu_03_1234} noted 31.4\% regulate other
transcription factors.

As the proportion of regulators of transcription factors rises,
the probable length of regulatory cascades increases. In fact, the
proportion of regulators taking part in a regulatory cascade of
length $n\geq 1$ is
\begin{equation}
   p_n = (1-P_{rr}) P_{rr}^{n-1}.
\end{equation}
This equation can be obtained from a tree of all binary pathways
which at each branching point either terminate with probability
$(1-P_{rr})$ or cascade with probability $P_{rr}$.  As such, the
probable cascade length is negligible when the proportion of
regulators controlling regulators is small $P_{rr}\ll 1$ but can
become large as $P_{rr}$ itself increases.  The calculated lengths
of regulatory cascades can be compared to those in {\em E. coli}
where 37.7\% are of length two, 52.5\% are three-level cascades,
and 9.8\% are four-level cascades \cite{MandanBabu_03_1234}. As
one-level or autoregulatory interactions are not included in this
observation, the predicted proportions for {\em E. coli} are
$\bar{p}_n=p_n/(1-p_1)$ with $P_{rr}=19.6\%$ giving 80\% two-level
cascades, 16\% three-level cascades, 3\% four-level cascades, 1\%
five-level cascades, and so on.  It is seen that the theoretical
predictions overestimate the proportion of two-level cascades and
underestimate the number of three-level cascades probably because
of selection pressures not included in the model, while other
calculated values closely approximate those observed.

We note that this model is entirely unable to explain the high
proportion of autoregulation observed in {\em E. coli} with
various estimates that 28.1\% \cite{Rosenfeld_02_785}, 50\%
\cite{Shen_Orr_02_64} and 46.9\% \cite{Thieffry_98_43} of
regulators are autoregulatory.  The predicted proportion of
autoregulators is approximated by replacing the very last fraction
($R/N$) in Eq. \ref{eq_regulation_regulators} by the term $1/N$
giving the probability that a self-directed link is formed,
leading to the expected autoregulatory proportion $\approx
l/(cg_oN)\approx 0.08\%$ for {\em E. coli}.  This failure likely
reflects the action of selection processes promoting spatial
rearrangements of entire regulons on the genome and the internal
shuffling of genes and promotor units. Such reorganizations of
duplicated gene regions (presumably shuffling genes and promotor
regions) have been common in {\em E. coli} allowing for instance,
spatial regulatory motifs whereby the promotors of colocated
(overlapping) and often co-functional operons transcribed in
opposing directions can interfere \cite{Warren_03_0310029}.

We now turn to consider the size of the largest connected island
in growing prokaryote gene networks featuring directed links whose
tails are preferentially attached to regulators and whose heads
are randomly distributed over all existing nodes.  For simplicity,
we define an island to consist of all nodes which are linked
regardless of the orientation of all links and so effectively
treat links as being undirected. This is because a regulator can
potentially perturb every node downstream to it including those
nodes downstream of other regulators and so can modify the
regulatory effects of other regulators---essentially, if the
downstream effects of different regulators eventually intersect,
we count these regulators in the same island. (Other definitions
of islands could be used.)

\begin{figure}[htb]
\centering
\includegraphics[width=\columnwidth,clip]{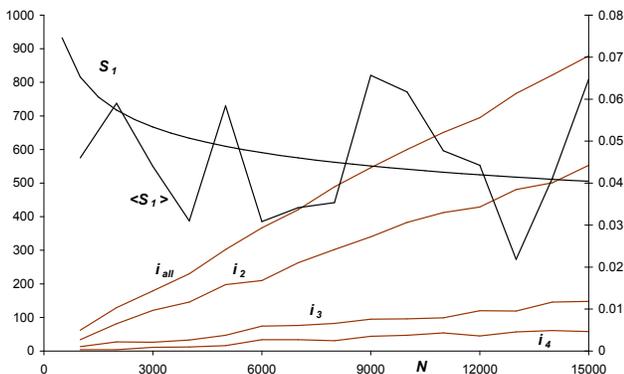}
\caption{\em The total number of discrete disconnected islands
$i_{\rm all}$, the number of islands with respectively two
($i_2$), three ($i_3$) and four ($i_4$) members (left hand axis),
and the simulated ($\langle s_1\rangle$) and predicted ($s_1$)
size of the largest island measured as a proportion of nodes for
various genome sizes (right hand axis).
}\label{f_e_coli_constant_p1_island}
\end{figure}

The growth of the largest island can be both directly simulated
and calculated under the continuum approximation
\cite{Gagen_0312021} (though this simple approach is indicative
only and is quite sensitive to for instance, the assumed average
size of external islands). The dominant (but not sole) mechanism
by which island $s_1$ can grow is for the newly added node $n_k$
to either (a) be a regulator (with probability $[1-(1-p)^m]=cg_o$)
and establish an outbound regulatory link to some existing node in
$s_1$ (with probability $s_1/k$) while at the same time accepting
a regulatory link (with probability $cg_o$) from a node in a
different island $s_{j\neq 1}$ (with probability $(k-s_1)/k$), or
(b) accept an inbound regulatory link (with probability $cg_o$)
from a regulator in island $s_1$ (with probability $s_1/k$) while
establishing a regulatory link (with probability $cg_o$) to some
node in a different island $s_{j\neq 1}$ (with probability
$(k-s_1)/k$). (Here, we assume that regulators are uniformly
distributed over islands and the number of links within an island
scales with the size of the island to crudely model preferential
attachment.) The result is that island $s_1$ grows by the size of
the second island $\langle s_{j\neq 1}\rangle$. Altogether, the
rate of growth in the size of island $s_1$ is then
\begin{equation}          \label{eq_island_mI}
  \frac{ds_1}{dk} = 2 (cg_o)^2
     \frac{s_1 [k-s_1]}{k^2} \langle s_{j\neq 1} \rangle.
\end{equation}
For initial conditions, we assume that a first link appears when
the genome has $1/cg_0\approx 11$ nodes ($s_1(11)=2$).  A
consistent solution for this equation appears with island size
growing linearly with genome size, $s_1=ak$, with
$a=1-1/[2(cg_o)^2\langle s_{j\neq 1} \rangle]$ under the
assumption that sufficient small islands are created to ensure
$\langle s_{j\neq 1} \rangle$ remains a constant. Simulations show
the average size of outside islands to be very closely $\langle
s_{j\neq 1}\rangle=3.31$ over a large range of genome sizes,
though a reasonable match between theory and simulation requires
setting $\langle s_{j\neq 1}\rangle=50$.  This is reasonable given
the approximations made. Fig. \ref{f_e_coli_constant_p1_island}
shows the number of all discrete islands as well as the number of
islands containing two, three and four components, and the
predicted and simulated sizes of the largest island expressed as a
proportion of the total genome size with a close match between
theory and simulation. This figure suggests that the {\em E. coli}
genome of $N=2528$ operons should possess a giant component
containing about 4\% of all nodes or about 100 operons.  This can
be compared to the observed figure where about 300 operons of the
examined regulatory and regulated operons (but not including
unregulated and nonregulatory operons) can be loosely grouped into
3-6 ``dense overlapping regulons" or DORS of about 50 operons each
while the remaining operons appeared as disjoint systems with most
containing 1-3 operons but some containing up to 25 operons
\cite{Shen_Orr_02_64}. The constant proportion of the genome taken
up by the largest island, and the constantly growing number of
discrete islands means that this network architecture suffers no
maximum size limit.  As a result, this approach is unable to
explain the upper size limit observed in the evolutionary record.

\subsection{Two-parameter nonaccelerating prokaryote network
model} \label{sect_non_accel_model_II}

The above one-parameter model combined the probability of forming
a link $p$ and the maximum number of links $m$ to give the
probable number of regulators formed $cg_o$. A two-parameter
nonaccelerating model can be constructed by delinking these
variables so that the probability of being a regulator is given
directly by $p\rightarrow r=cg_o$ leaving the number of links
established $m$ as a free parameter.  This gives the number of
regulators as $R=cg_oN=rN$.  We assume that every regulator $n_k$
gains exactly $t_{kk}=m$ outbound links on entry to the genome
which are randomly distributed as inbound link heads over all
existing nodes.  The average number of initial outbound links per
regulator is then $m$ while the average number of outbound links
per node is $rm$. These outbound link tails must be balanced by
uniformly distributed inbound link heads so consequently, we
assume that all nodes on entry to the genome receive inbound
regulatory links distributed according to
\begin{equation}
 P(j) = {m \choose j} r^j (1-r)^{m-j},
\end{equation}
giving the required average of inbound links per node of $\langle
h_{kk}\rangle=rm$. The average number of links is $L=2rmN$ taking
account of both heads and tails.  Hence, the number of outbound
links per regulator is $L/R\approx 2m$, so setting $m=2$ allows a
close fit between this model and the value of $5$ observed in {\em
E. coli}. This sets $L=2rmN=0.374N$ giving the number of inbound
links per node as $L/N=2rm=0.374$. The values of the link
formation probability $r$ suggest an overly short average promotor
binding site length of $-\log_4 r=1.71$ bases.

\subsubsection{Random distribution of regulated operons---II}

With respect to the distribution of inbound regulatory links, the
two-parameter model does not differ in any material respect from
the earlier one-parameter model as in both cases links are
uniformly distributed over all nodes. However, the link formation
probability differs in each approach, so all of the results of
Eqs. \ref{eq_continuum_heads} to \ref{eq_unreg_operons} can be
used with the changes $p\rightarrow r=cg_o$, $m=20\rightarrow
m=2$, and $l=0.196\rightarrow 2rm=0.374$.

\begin{figure}[htb]
\centering
\includegraphics[width=\columnwidth,clip]{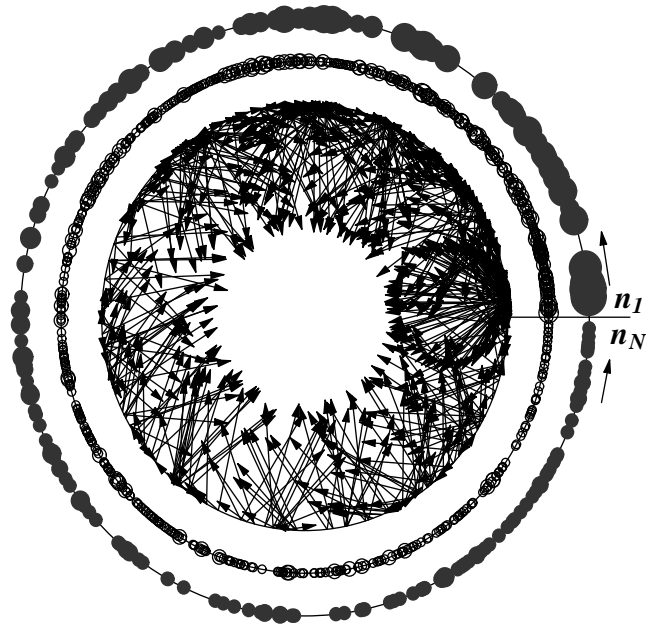}
\caption{\em An example statistically generated E. coli genome
using a two-parameter constant growth model using the same
settings as in Fig. \protect\ref{f_e_coli_constant_p1}. Note that
the distributions of regulators and of regulated operons over the
genome are uncorrelated with age, while the numbers of inbound and
outbound regulatory links are strongly correlated with age.}
\label{f_e_coli_constant_p2}
\end{figure}

Consequently, the distribution of inbound regulatory heads over
all nodes is
\begin{equation}
  h_{jN} = rm \left[1+\ln\left(\frac{N}{j} \right)\right],
\end{equation}
again monotonically decreasing with node age. This distribution
suggests that the oldest node $n_1$ for the {\em E. coli} genome
with $N=2528$ nodes will possess $h_{1N}=1.65$ inbound regulatory
links while the most recent node $h_{NN}$ will possess
$h_{NN}=0.187$ inbound regulatory links---see the age dependent
distributions of Fig. \ref{f_e_coli_constant_p2}.

\begin{figure}[htbp]
\centering
\includegraphics[width=\columnwidth,clip]{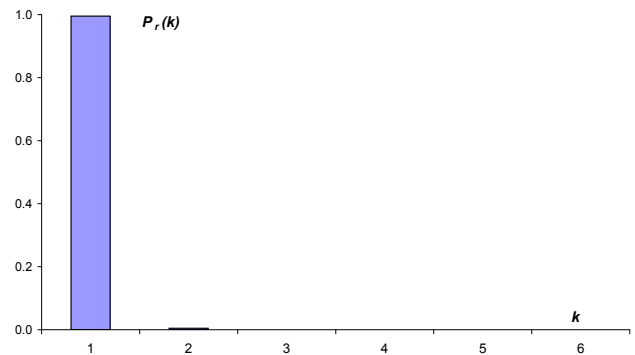}
\caption{\em The predicted proportions $P_r(k)$ of the regulated
operons of E. coli taking multiple regulatory inputs for genomes
of any size. This distribution poorly approximates that observed
for E. coli with $N=2528$ operons in Fig. 2(d) of Ref.
\protect\cite{Cherry_03_40} and of Fig. 5 of Ref.
\protect\cite{MandanBabu_03_1234}.
\label{f_heads_per_reg_operon_II}}
\end{figure}

Following the previous derivation, the distribution of inbound
link numbers for regulated nodes (i.e. those with $k>0$) is
\begin{equation}            \label{eq_reg_link_number_II}
  P_r(k) =  \left( e^{1/rm}-1 \right) e^{-k/rm},
\end{equation}
which again is normalized to unity. As previously, the number of
regulated nodes is approximately $O_r=N-O_u\approx 2rmN$, so in
turn, the number of inbound links per regulated node is $L/O_r=1$.
A direct calculation of the average number of inbound links for
regulated operons using the distribution of Eq.
\ref{eq_reg_link_number_II} gives $\langle
k\rangle=\sum_{k=1}^{\infty} kP_r(k)=1/[1-\exp(-1/rm)=1.0047$,
close to the value of $1.5$ or 1.6 observed in {\em E. coli}
\cite{Thieffry_98_43,Shen_Orr_02_64,MandanBabu_03_1234}.  The
predicted distribution of inbound links for regulated operons (Eq.
\ref{eq_reg_link_number_II}) can be compared to that observed in
the {\em E. coli} network of size $N=2528$ operons
\cite{Cherry_03_40}, and is shown in Fig.
\ref{f_heads_per_reg_operon_II}. Again, the overly rapid decay of
the calculated distribution poorly approximates the compact
exponential distribution observed for {\em E. coli} shown in Fig.
2(d) of Ref. \cite{Cherry_03_40} and of Fig. 5 of Ref.
\cite{MandanBabu_03_1234} leading to an underestimation of the
numbers of regulated operons with 3 or more inputs---essentially
no regulators are predicted to have 3 or more inputs for genomes
of size $N=2528$ operons.

\subsubsection{Scale-free distribution of regulator operons---II}

As previously, the rate of growth in outbound link number for node
$n_j$ is approximately
\begin{equation}
  \frac{\partial t_{jk}}{\partial k} =
     h_{kk}\frac{t_{jk}}{2rmk}.
\end{equation}
Here, the denominator on the right hand side is the expected
number of existing links in a network of size $k$ nodes. Noting
initial conditions $t_{jj}=rm$, and $h_{kk}=rm$, we have
\begin{equation}
  t_{jN} = rm \left( \frac{N}{j} \right)^{1/2}.
\end{equation}
As previously, this is the density of outbound regulatory links
per node which equates to the density of outbound regulatory links
per regulator times the density of regulators per node.  As this
latter density is uniform over the genome and equal to $R/N=r$,
then the density of outbound links per regulator is
\begin{equation}         \label{eq_t_ii_dist_II}
  t_r(j,N) = m \left( \frac{N}{j} \right)^{1/2}.
\end{equation}
Again, this distribution is monotonically decreasing with node age
so older nodes are more heavily connected---see Fig.
\ref{f_e_coli_constant_p2}.  With the additional degree of freedom
offered by the independent parameter $m$, this distribution shows
the most heavily connected regulators having around 100 links in
{\em E. coli} (with $j=1$, $m=2$, and $N=2528$).

The often unknown age information in the $t_r(j,N)$ distribution
can be discarded using the continuum approach (noting
$k=m(N/j)^{(1/2)}$) to obtain the outbound link distribution
\begin{equation}   \label{eq_k_minus_three_II}
   T(k,N) =  \frac{2 m^2}{k^3},
\end{equation}
which is normalized over the range $[m,\infty)$ as
$\int_{m}^\infty T(k,N)=1$. In turn, the expected proportion of
regulators $P(k)$ possessing $k$ links is then obtained by
integrating the continuous distribution of Eq.
\ref{eq_k_minus_three_II} over appropriate ranges $[m,5/2]$ or
$[k-1/2,k+1/2]$ to obtain
\begin{equation}
    P_t(k) = \left\{
      \begin{array}{ll}
        1-\left( \frac{2m}{5} \right)^2, & k=2 \\
         &  \\
        32 m^2 \frac{k}{(4k^2-1)^2}, & k>2. \\
      \end{array}
    \right.
\end{equation}
Here, $k\geq 2$ as the choice $m=2$ means that the minimum number
of links that a regulator can possess is two. As required, this is
normalized to unity. The average number of outbound links per
regulator is, using the continuous distribution $T(k,N)$, $\langle
k\rangle=\int_{m}^\infty kT(k,N)=2m=4$  and numerically calculated
to be $\langle k\rangle=3.9$ using the $P_t(k)$ distribution each
of which compares well to the observed value of 5 in {\em E. coli}
\cite{Shen_Orr_02_64}.

\begin{figure}[htb]
\centering
\includegraphics[width=\columnwidth,clip]{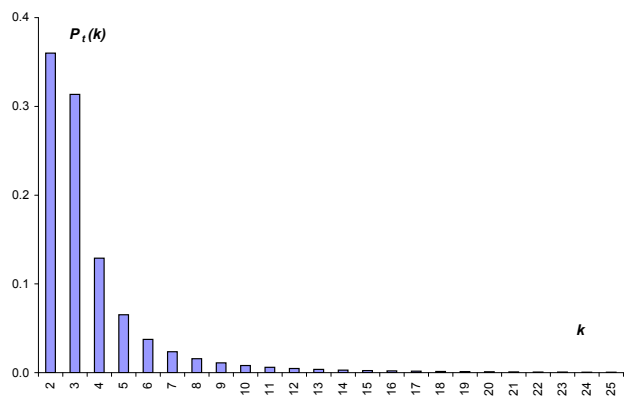}
\caption{\em The predicted proportion of regulatory operons
$P_t(k)$ regulating $k$ different operons for arbitrary gene
networks. As expected, most regulators regulate only one other
operon, though a small number of regulators can regulate more than
50 operons. This distribution poorly approximates the observed
proportions for E. coli in Fig. 2(c) of Ref.
\protect\cite{Cherry_03_40} and Fig. 4 of Ref.
\protect\cite{MandanBabu_03_1234}, and predicts the probable
existence of one E. coli regulator possessing link numbers in each
of the respective ranges between $[70,99]$ links, and between
$[100,\infty)$ links. \label{f_e_coli_long_tails_II}}
\end{figure}

Again, the very rapid (cubic) decrease in probable link numbers
means that these distributions have difficulty in reproducing the
distributions observed in {\em E. coli}
\cite{Cherry_03_40,MandanBabu_03_1234}. The expected outbound link
distribution appears in Fig. \ref{f_e_coli_long_tails_II} showing
that 36\% of regulators have two links, while 67\% have three or
fewer links, and 80\% have four or fewer links. In particular, the
expected number of regulators with $k$ links is $P_t(k)R$ with
$R=rN$. For {\em E. coli} with $N=2528$ operons
\cite{Cherry_03_40}, this predicts the probable existence of only
one {\em E. coli} regulator possessing link numbers in each of the
respective ranges between $[70,99]$ links and in the range
$[100,\infty)$ links. This poorly approximates the connectivity of
{\em E. coli} where many regulators regulate more than 20 operons
including the global food sensor CRP which regulates up to 197
genes directly \cite{MandanBabu_03_1234}.

\begin{figure}[htb]
\centering
\includegraphics[width=\columnwidth,clip]{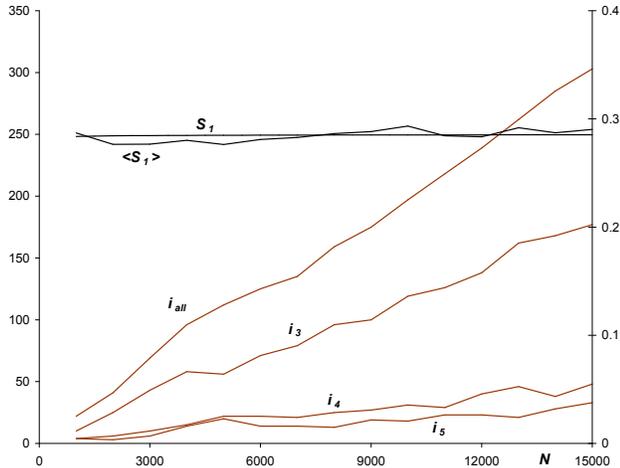}
\caption{\em The total number of discrete disconnected islands
$i_{\rm all}$, the number of islands with respectively three
($i_3$), four ($i_4$) and five ($i_5$) members (left hand axis),
and the simulated ($\langle s_1\rangle$) and predicted ($s_1$)
size of the largest island measured as a proportion of nodes for
various genome sizes (right hand axis). We note that the choice
$m=2$ ensures there are no two member islands.}
\label{f_islands_constant_p2}
\end{figure}

\begin{table*}[htb]
\begin{ruledtabular}
\begin{tabular}{l|lllll}
 $m \backslash N$ &  3,000      &  6,000      &  9,000      &   12,000     &  15,000        \\ \hline
 1                &  0.061 (85) &  0.011 (346)&  0.019 (531)&   0.016 (706)&  0.013 (888)   \\
 2                &  0.277 (69) &  0.281 (125)&  0.288 (175)&   0.283 (239)&  0.290 (303)   \\
 3                &  0.447 (21) &  0.436 (32) &  0.448 (41) &   0.453 (52) &  0.450 (67)    \\
 4                &  0.554 (4)  &  0.542 (4)  &  0.549 (7)  &   0.550 (9)  &  0.548 (11)    \\
 5                &  0.622 (1)  &  0.610 (2)  &  0.618 (2)  &   0.620 (1)  &  0.619 (2)     \\
\end{tabular}
\end{ruledtabular}
\caption{\label{t m_settings}  The relative size of the largest
island component and the total number of islands (in brackets) for
networks of various sizes $N$ and for different choices for the
initial number of links per node or regulator $m$.}
\end{table*}

\subsubsection{Cascades and regulatory islands---II}

The proportion of transcription factors which control downstream
regulators is
\begin{eqnarray}    \label{eq_regulation_regulators_II}
   P_{rr}(N) &=& \frac{1}{R} \sum_{k=1}^N
     r
     \left[ m \left( \frac{N}{k}\right)^{\frac{1}{2}}\right]
     \frac{R}{N} \nonumber  \\
     &\approx & 2rm.
\end{eqnarray}
Here, the derivation follows that of Eq.
\ref{eq_regulation_regulators}. Again, the proportion of
regulators which control transcription factors is independent of
network size and equals 37.4\%, which compares well with the
31.4\% observed in {\em E. coli} \cite{MandanBabu_03_1234}.

As previously, the proportion of regulators taking part in a
regulatory cascade of length $n$ is $\bar{p}_n=p_n/(1-p_1)$ with
$P_{rr}=37.4\%$ giving 63\% two-level cascades, 23\% three-level
cascades, 9\% four-level cascades, 3\% five-level cascades, 1\%
six-level cascades, and so on. As previously, these ratios
overestimate the proportion of two-level cascades and
under-estimate the proportion of higher level cascades in {\em E.
coli} with 37.7\% two-level cascades, 52.5\% three-level cascades,
9.8\% four-level cascades \cite{MandanBabu_03_1234}.

The size of the largest connected island is again expected to
occupy a constant proportion of the genome regardless of size. An
equivalent derivation to that of Eq. \ref{eq_island_mI} gives the
rate of growth in the size of island $s_1$ as
\begin{equation}
  \frac{ds_1}{dk} = 2 (rm)^2
     \frac{s_1 [k-s_1]}{k^2} \langle s_{j\neq 1} \rangle.
\end{equation}
For initial conditions, we assume that a first link appears when
the genome has $1/r\approx 11$ nodes giving $s_1(11)=3$ as the
choice $m=2$ ensures there are no two member islands.  As
previously, a consistent solution exists with island size growing
linearly with genome size ($s_1=ak$) with $a=1-1/[2(rm)^2\langle
s_{j\neq 1} \rangle]$ under the assumption that sufficient small
islands are created to ensure $\langle s_{j\neq 1} \rangle$
remains a constant. Simulations show the average size of outside
islands to be very closely $\langle s_{j\neq 1}\rangle=4.17$ over
a large range of genome sizes, while a reasonable match between
theory and simulation requires setting $\langle s_{j\neq
1}\rangle=20$, which is reasonable given the approximations made.
Fig. \ref{f_islands_constant_p2} shows the number of all discrete
islands as well as the number of islands containing three, four,
and five components, and the predicted and simulated sizes of the
largest island expressed as a proportion of the total genome size
with a close match between theory and simulation. This figure
suggests that the {\em E. coli} genome of $N=2528$ operons should
possess a giant component containing about 30\% of all nodes or
about 460 operons which overestimates that observed
\cite{Shen_Orr_02_64}. Again, the constant proportion of the
genome taken up by the largest island, and the constantly growing
number of discrete islands means that this network architecture
suffers no maximum size limit.  As a result, this approach is
unable to explain the upper size limit observed in the
evolutionary record.

The two-parameter model has been developed with the setting $m=2$
to best match the observed number of links per regulator. However,
a setting $m=3$ provides at least as good a match, and it is
possible that choosing alternate settings for the initial number
of links per regulator ($m$) and per node ($rm$) might improve the
fit to the data.  Table \ref{t m_settings} shows the relative size
of the largest island and the total number of islands for
simulated genomes of different size and for different choices of
$m$ It is clear that choices $m\geq 3$ overestimates the size of
the largest regulatory islands while the choice $m=1$ gives a poor
fit to the observed number of regulatory links per regulator.

\section{Conclusion}

In this paper, we developed two probabilistic nonaccelerating
network models for the growth of prokaryote regulatory gene
networks. These models complement the accelerating network model
presented in Ref. \cite{Gagen_0312021} allowing a comparison of
these alternate approaches.

Each of the nonaccelerating models presented here faces
considerable difficulties in providing a plausible physical
mechanism justifying a nonaccelerating regulatory model, and fails
to consider any additional steric or logical limitations on
combinatoric control at any given promotor. Further, these
approaches are unable to explain the observed quadratic growth in
prokaryote regulator number with increasing genome size displayed
in Fig. \ref{f_prok_reg_data}. This mismatch between predicted and
observed numbers of regulators is also reflected in the overly
short expected promotor sequence lengths in each model. Further,
the linear growth in regulator number with genome size effectively
means that these networks are becoming relatively more and more
sparsely connected with growth---the desired maximum number of
possible links grows as $N^2$ so the relative density of links
goes as $L/N^2\propto 1/N\rightarrow 0$ as $N$ becomes large. This
decrease in relative connection density means that nonaccelerating
networks suffer their own inherent size constraints as complex
networks operate poorly when sparsely connected.

We further compared each model to observed results for {\em E.
coli}, and achieved reasonable matches for the average
connectivity of the long tailed distribution of outgoing
regulatory links (approximately 5) and the average of the
exponential distribution of incoming regulatory links
(approximately 1.5).  However, the distributions themselves were
either overly lightly connected (model one) or decayed overly
rapidly leading to a distinct under-representation of highly
connected nodes compared to the {\em E. coli} distributions
(models one and two). Each of the nonaccelerating models was able
to reasonably match the observed proportion of regulators
controlling regulators (approximately 31.4\%) and in turn, the
probable length of regulatory cascades. Lastly, the first of the
nonaccelerating models was able to roughly reproduce {\em E. coli}
statistics on the numbers of discrete regulatory islands, though
the second model overestimated the size of the largest discrete
regulatory island. Because of the size independent statistics of
these nonaccelerating models, neither approach displays structural
transitions at any critical network size and thus face
difficulties in explaining the prokaryote size and complexity
limitations evident in the evolutionary record.

Our approach in this paper (and in Ref. \cite{Gagen_0312021}) is
unable to explain the high proportion of autoregulation observed
in {\em E. coli} \cite{Shen_Orr_02_64} and this failure likely
points to selection for genome reorganizations leading to spatial
arrangements of operons allowing joint regulation
\cite{Warren_03_0310029} which is not included in this model.
Further, this approach does not include selection pressures
ensuring that similarly regulated islands or modules share common
functionality \cite{Shen_Orr_02_64}, or other regulatory
mechanisms influencing both the transcription and translation of
transcription factors including micro-RNAs and other chemical
mechanisms and mediators (see for instance \cite{Vogel_03_6435}).

The accelerating and nonaccelerating models of prokaryote gene
networks differ most markedly in their predictions for the age
dependency of the distribution of inbound and outbound regulatory
links.  It would be interesting to obtain information on the
correlation (if any) between age and link number for different
prokaryotes to properly distinguish these approaches.

We conclude that viable models of prokaryote regulatory gene
networks are likely to be accelerating in nature. This is
important as much current network analysis is predicated on the
assumption that only nonaccelerating networks are relevant to
society or biology due to their unconstrained sizes and constant
statistics. However, such assumptions make it very difficult to
explain the size limitations displayed by prokaryotic gene
networks in the evolutionary record. Subsequently, it is likely
that viable models of eukaryotic regulatory networks will be
accelerating and will incorporate computationally complex
technologies \cite{Mattick_01_1611,Mattick_01_986,Mattick_03_930}.


\end{document}